\documentclass[aps,prd,twocolumn,superscriptaddress,floatfix,nofootinbib,natbib]{revtex4-1}
\usepackage{graphicx,longtable,mathrsfs,color,array}
\usepackage[usenames,dvipsnames]{xcolor} 
\usepackage{amssymb,amsmath,mathtools,mathrsfs,slashed} 
\usepackage{epsfig,subfigure,placeins,float} 
\usepackage{booktabs,longtable,ctable,multirow} 
\usepackage{exscale,relsize} 
\usepackage[normalem]{ulem} 
\usepackage{enumerate}
\usepackage[colorlinks,linkcolor=blue,citecolor=blue,urlcolor=blue ]{hyperref}

\usepackage[stable]{footmisc}
\usepackage{color}

\begin{document}
\title{Time-varying neutrino mass from a supercooled phase transition: current cosmological constraints and impact on the $\boldsymbol{\Omega_m}$-$\boldsymbol{\sigma_8}$ plane}
\author{Christiane S. Lorenz}
\email{christiane.lorenz@physics.ox.ac.uk}
\affiliation{Astrophysics, University of Oxford, DWB, Keble Road, Oxford OX1 3RH, UK}
\author{Lena Funcke}
\email{lfuncke@perimeterinstitute.ca}
\affiliation{Max-Planck-Institut f{\"u}r Physik (Werner-Heisenberg-Institut), F{\"o}hringer Ring 6, 80805 M{\"u}nchen, Germany}
\affiliation{Arnold Sommerfeld Center, Ludwig-Maximilians-Universit{\"a}t, Theresienstra{\ss}e 37, 80333 M{\"u}nchen, Germany}
\affiliation{Perimeter Institute for Theoretical Physics, 31 Caroline Street North, Waterloo, ON, N2L 2Y5, Canada}
\author{Erminia Calabrese}
\affiliation{School of Physics and Astronomy, Cardiff University, The Parade, Cardiff, CF24 3AA, UK}
\author{Steen Hannestad}
\affiliation{Department of Physics and Astronomy, Aarhus University, Ny Munkegade 120, DK–8000 Aarhus C, Denmark}

\date{Received \today; published -- 00, 0000}

\begin{abstract}
In this paper we investigate a time-varying neutrino mass model, motivated by the mild tension between cosmic microwave background (CMB) measurements of the matter fluctuations and those obtained from low-redshift data.
We modify the minimal case of the model proposed in Ref.~\cite{Dvali:2016uhn} that predicts late neutrino mass generation in a post-recombination cosmic phase transition, by assuming that neutrino asymmetries allow for the presence of relic neutrinos in the late-time Universe. We show that, if the transition is supercooled, current cosmological data (including CMB temperature, polarization and lensing, baryon acoustic oscillations, and Type Ia supernovae) prefer the scale factor $a_s$ of the phase transition to be very large, peaking at $a_s\sim 1$, and therefore supporting a cosmological scenario in which neutrinos are almost massless until very recent times. We find that in this scenario the cosmological bound on the total sum of the neutrino masses today is significantly weakened compared to the standard case of constant-mass neutrinos, with $\sum m_\nu<4.8$~eV at 95\% confidence, and in agreement with the model predictions. The main reason for this weaker bound is a large correlation arising between the dark energy and neutrino components in the presence of false vacuum energy that converts into the non-zero neutrino masses after the transition. This result provides new targets for the coming KATRIN and PTOLEMY experiments. We also show that the time-varying neutrino mass model considered here does not provide a clear explanation to the existing cosmological $\Omega_m$-$\sigma_8$ discrepancies.
\end{abstract}


\maketitle

\section{Introduction}
The absolute value and the origin of the neutrino masses are two of the main open questions in particle physics and cosmology. The discovery of neutrino oscillations~\cite{PhysRevLett.81.1562,PhysRevLett.87.071301,PhysRevLett.89.011301} implies that at least two of the three neutrino mass eigenstates must have a non-vanishing mass, and gives a lower limit of $59$ meV (normal ordering) and $109$ meV (inverted ordering) for the total sum of the neutrino masses, $\sum m_\nu$~\cite{Tanabashi}. For the high-end tail of the mass distribution, the most stringent upper limit is set by cosmological data. Observations of the cosmic microwave background (CMB) from the \textit{Planck} satellite, combined with baryon acoustic oscillations (BAO), give~$\sum m_\nu<0.12$ eV at 95\% confidence~\cite{Aghanim:2018eyx}. The projected sensitivity of future CMB and BAO data is $\sim 0.03$~eV~\cite{CMB-S4, SO}. This is an indirect measurement tracking the effect of the neutrino masses on the matter distribution in the Universe. Upper limits on the absolute electron neutrino mass have also been obtained from direct $\beta$-decay searches (see e.g.\ Refs.~\cite{Lobashev2003,Kraus2005,Aseev2011}), with $m_{\nu_e}\leq 2.2$~eV at 95\% confidence. The KATRIN $\beta$-decay experiment will improve these limits by measuring $m_{\nu_e}$ down to $0.2$ eV at 90\% confidence~\cite{Drexlin:2013lha}; this is about one order of magnitude higher than future cosmological sensitivity. Flavour eigenstates  of  the  neutrino, such as the electron neutrino mentioned here, can be described as  linear  combinations of the neutrino mass eigenstates and are connected to those through the Pontecorvo-Maki-Nakagawa-Sakata mixing matrix~\cite{doi:10.1143/PTP.28.870,Pontecorvo:1967fh,Tanabashi}. Cosmology and laboratory searches are sensitive to different linear combinations of the neutrino mass eigenstates and therefore confine the neutrino parameter space in a complementary way. 

The discovery of neutrino oscillations hints at fundamental new physics beyond the Standard Model (SM) of particle physics, since the SM particle content does not allow for any renormalizable neutrino mass term \cite{Schwartz2013}. In fact, Dirac neutrino masses cannot be accommodated in the SM due to the absence of right-handed neutrino states, and Majorana masses for the left-handed neutrinos are not allowed as the SM Higgs sector only contains an $SU(2)_L$ doublet and no triplets. Therefore, it is widely believed that neutrino masses require the postulation of new elementary particles (see e.g.\ Refs.\ \cite{Tanabashi,deGouvea2016, Hernandez2017} for reviews). The most popular directions of model building beyond the SM usually focus on new physics at short distances corresponding to high energies ($E\gtrsim$ TeV), and thereby strongly affecting early-Universe cosmology.
As an alternative direction, Ref.~\cite{Dvali:2016uhn} proposed a low-energy solution to the neutrino mass problem at a new infrared gravitational scale ($\Lambda_G\lesssim$ eV), which is numerically coincident with the scale of dark energy. As reviewed below, this model alters late-Universe cosmology after photon decoupling.

In the Standard Model of cosmology, and/or in the presence of these relic neutrinos with time-varying mass, the neutrino mass affects the growth of cosmic structures in several ways (see e.g.\ Ref.~\cite{lesgourgues_mangano_miele_pastor_2013} for a review and Ref.~\cite{Lorenz:2017fgo} for a summary of the effects relevant here). In particular, non-zero masses suppress the amplitude of matter fluctuations in the late-time Universe compared to those present at early times, i.e.\ at the time of the CMB decoupling. Therefore, the total sum of the neutrino masses is strongly correlated with the inferred values of the matter density, $\Omega_m$, and matter clustering, for example measured by the amplitude of matter fluctuations on 8 $h^{-1}$ Mpc scales, $\sigma_8$. These quantities can be constrained with the CMB, the CMB lensing signal (that is the deflection of the CMB photon paths due to gravitational potential wells along their trajectories), and different probes of the matter distribution in the local Universe, e.g.\ the galaxy weak lensing signal, galaxy clustering, and the abundance of galaxy clusters. 

Over the past few years, measurements of $\Omega_m$-$\sigma_8$ from early- and late-time surveys have shown some mild tensions. In particular, taking the parameter combination $S_8\equiv\sigma_8\sqrt{\Omega_m/0.3}$, the tension exists when comparing \textit{Planck} CMB constraints~\cite{planck2018params} with galaxy weak lensing data from the Canada France Hawaii Lensing Survey (CFHTLenS) at the 1.7$\sigma$ level~\cite{MacCrann:2014wfa} (see also Ref.~\cite{Joudaki:2016mvz}), from the Kilo Degree Survey (KiDS) at the 2.2$\sigma$ level~\cite{Hildebrandt:2016iqg,Joudaki:2016kym} (2.6$\sigma$ in combination with 2dFLenS~\cite{Joudaki:2017zdt}), and from the first-year release of the Dark Energy Survey (DES) at the 1.7$\sigma$ level~\cite{Troxel:2017xyo}. Similar levels of inconsistency are found between $\Omega_m$-$\sigma_8$ inferred from the abundance of galaxy clusters detected with the Sunyaev--Zel'dovich (SZ) effect and \textit{Planck} CMB values~\cite{Ade:2013lmv,Ade:2015fva}.
This has generated a lot of interest in the cosmological community with efforts split between investigation of residual systematics in the data or analysis assumptions in KiDS, DES, and \textit{Planck}~(e.g.~\cite{Addison:2015wyg,Efstathiou:2017rgv,Sellentin:2017koa,Troxel:2018qll,Obied:2017tpd,Asgari:2018hde}), and the possibility of having seen signatures of new physics beyond the standard $\Lambda$CDM cosmological model~(e.g.~\cite{Barros:2018efl,McCarthy:2017csu,Joudaki:2016kym,Poulin:2018zxs,Peirone:2017lgi,Lesgourgues:2015wza,Poulin:2016nat,Buen-Abad:2017gxg,DiValentino:2015bja,Gomez-Valent:2017idt}). 
For example, Refs.~\cite{Joudaki:2016kym,Battye:2013xqa,Wyman:2013lza} explored whether time-varying dark energy or neutrino masses might solve the tensions. Although the significance of the discrepancy changes slightly in more extended models, there is, at present, no clear preference for a beyond-$\Lambda$CDM cosmology. 

However, a general trend of these results is that low-redshift data prefer less matter fluctuations compared to early-time estimates which, when allowing neutrino masses to vary, translates into higher preferred values of the neutrino mass compared to the constraints coming from the CMB alone. Motivated by this, and taking at face value the analysis assumptions and the likelihood packages of each experiment (i.e.\ assuming this is not data/analysis systematics driven), we explore here a time-varying neutrino mass model, where the neutrino mass increases with time. Time-varying neutrino mass models were first introduced by Ref.~\cite{Fardon:2003eh} as a way to explain the similar energy scales of massive neutrinos and dark energy, and suggested that mass-varying neutrinos could be the cause of cosmic acceleration. However, Ref.~\cite{Afshordi:2005ym} showed that these models would not be stable, and not distinguishable from a cosmological constant. Time-varying neutrino mass models and their cosmological implications were also studied in Refs.~\cite{Franca:2009xp,2013A&A...560A..53L,Brookfield:2005td,Brookfield:2005bz,Bjaelde:2007ki,Geng:2015haa}.

A new time-varying neutrino mass model was recently proposed in Ref.~\cite{Dvali:2016uhn}, where neutrino masses are generated through a gravitational $\theta$-term in a late cosmic phase transition. This transition is expected to be of first order (see anologous discussions in Ref.~\cite{Pisarski:1983ms,Roder:2003uz}) and thus can either take place almost instantaneously at a temperature $T\sim m_\nu$ or can be substantially supercooled and thus become apparent only at lower temperatures $T\lesssim m_\nu$. In both cases, the minimal case of the gravitational mass model predicts almost complete relic neutrino annihilation after the transition, so that all cosmological mass constraints are entirely evaded. However, a substantial relic neutrino density can survive in the non-minimal case of neutral lepton asymmetries, which was not considered in Ref.~\cite{Dvali:2016uhn}. In this case, impact on neutrino mass constraints from cosmological data would be expected. For example, cosmological constraints on a simplified version of this non-minimal case of the model were presented in Ref.~\cite{Koksbang:2017rux}, finding that in some cases the cosmological neutrino mass bounds are significantly weakened compared to the standard constant-mass case, with $\sum m_\nu \lesssim 0.6-0.8$~eV.

In this paper, we extend the analysis of Ref.~\cite{Koksbang:2017rux} in three ways. 1) We include false vacuum energy from the supercooled phase transition, which is especially important when generating relatively large neutrino masses at late times corresponding to low temperatures (see Section~\ref{sec:Theory}), this was neglected in Ref.~\cite{Koksbang:2017rux}. For simplicity, we neglect the neutrino self-interactions and (partial) annihilation, as predicted by the model in Ref.~\cite{Dvali:2016uhn}, which will be treated in future studies. 2) We add \textit{Planck} polarization data. 3) We examine whether time-varying neutrino masses can ease the tensions between cosmological parameters inferred from high- and low-redshift data, looking at the constraints from different probes in the $\Omega_m$-$\sigma_8$ plane.
The analysis assumptions are reported in Section~\ref{sec:analysis} and results in Section~\ref{sec:Tensions}. 
We summarize our findings and discuss the implications of our analysis both for the KATRIN experiment and for relic neutrino detection experiments, such as PTOLEMY~\cite{Betts:2013uya}, in Section \ref{sec:Discussions}.

\section{Time-varying neutrino mass model}\label{sec:Theory}

\subsection{Theoretical Foundations}

The gravitational neutrino mass model in Ref.~\cite{Dvali:2016uhn} predicts the relic neutrinos to be massless until a late cosmic phase transition after photon decoupling. In the transition, a neutrino vacuum condensate forms and generates small effective neutrino masses $m_\nu\sim\Lambda_G\sim|\langle\bar{\nu}\nu\rangle|\equiv v$, where $\Lambda_G$ is the neutrino flavor symmetry breaking scale and $v$ is the scale of the vacuum condensate\footnote{Ref.~\cite{Dvali:2016eay} showed that this scenario could also solve the strong $CP$ problem if the condensate generates the up-quark mass as well.}.
The massive relic neutrinos then rapidly decay into the lightest neutrino mass eigenstate, become strongly coupled, and (partially) bind up or annihilate into almost massless Goldstone bosons through the process $\nu + \bar{\nu}\rightarrow \phi + \phi$. Naively, one might expect this modification of the relic neutrino sector to be ruled out by cosmological observations; for example, the similar idea of a neutrinoless Universe \cite{Beacom2004} was ruled out by neutrino free-streaming in the early Universe \cite{Hannestad2004,Lancaster2017}, an induced phase shift in the CMB peaks~\cite{Follin:2015hya}, and precision measurements of the effective number of neutrino species from the CMB (more recently from Ref.~\cite{planck2018params}). This is not the case because, crucially, the temperature $T_{\Lambda_G}$ of the neutrino phase transition is a free parameter of the model in Ref.~\cite{Dvali:2016uhn}, fixed to $T_{\rm today}\lesssim T_{\Lambda_G}\lesssim T_{\rm CMB}$ by the above-mentioned cosmological constraints\footnote{We note that the upper bound on $T_{\Lambda_G}$ still applies if neutrinos get small masses through other mechanisms beyond gravity, making this constraint model-independent. A generic lower bound on the scale $\Lambda_G$ stems from experimental tests of Newtonian gravity down to $\sim {\rm meV}^{-1}$ distances \cite{Kapner2006}, which is similar to the model-dependent lower bound on $v$ from the observed neutrino mass splitting~\cite{Dvali:2016uhn}.}. Thus, Ref.~\cite{Dvali:2016uhn} predicts neutrino self-interactions and (partial) annihilation only in the late Universe after photon decoupling, making the model predictions cosmologically viable.

Additionally, an important point to stress here is that, although an almost complete relic neutrino annihilation is a key prediction of the minimal case in Ref.~\cite{Dvali:2016uhn}, it can be evaded in the presence of neutrino asymmetries. Big Bang Nucleosynthesis (BBN) and CMB data weakly constrain the muon- and tau-neutrino asymmetries \cite{planck2015xiii}, while BBN data strongly constrain the electron-neutrino asymmetry \cite{Mangano2011}. If standard neutrino oscillations in the early Universe mix the neutrino flavors, the strong BBN bounds would apply to all neutrino flavors \cite{Castorina2012}. However, the model in Ref.~\cite{Dvali:2016uhn} predicts massless relic neutrinos in the early Universe, and all flavor-violating couplings only turn on abruptly when approaching the late-time phase transition (similar to, e.g.\ axion couplings \cite{Sikivie2006}). 
We derive that the latest \textit{Planck} CMB limit of $\Delta N_{\rm eff}<0.33$ at 95\% confidence provides a weak bound on the $\nu_{\mu,\tau}$ asymmetries
\begin{equation}
\left|\frac{n_{\nu_{\mu,\tau}} - n_{\bar{{\nu}}_{\mu,\tau}}}{n_{\nu_{\mu,\tau}}}\right| \lesssim 0.16 \times \frac{11}{3} \sim 0.58\,,
\end{equation}
and therefore that up to $\sim 58\%$ of the $\nu_\mu$ and $\nu_\tau$ flavors could have survived the annihilation after the late phase transition. This corresponds to $\sim 39$\% of all relic neutrinos. Such an asymmetry could only survive in the Dirac neutrino case \cite{Langacker1982}, which implies that the Majorana case of Ref.~\cite{Dvali:2016uhn} would always yield a neutrinoless Universe. 

Consequently, in this work we consider a modified version of the minimal case in Ref.~\cite{Dvali:2016uhn}, exclusively studying late neutrino mass generation and neglecting the self-interactions and (partial) annihilation which we leave for future studies.

Ref.~\cite{Dvali:2016uhn} assumed the phase transition to take place instantaneously, i.e.\ at a temperature $T_{\Lambda_G}\sim \Lambda_G\sim v \sim m_\nu$. However, since the phase transition is expected to be of first order (see anologous discussions in Refs.~\cite{Pisarski:1983ms,Roder:2003uz}) and the possible presence of neutrino asymmetries would further strengthen the first-order transition (see analogous discussions in Refs.~\cite{Stephanov:1998dy,Boeckel2010}), the transition can also be substantially delayed and thus become \textit{apparent} only at lower temperatures. Such a supercooling mechanism is well known from inflationary and other cosmological scenarios (see e.g.\ Refs.~\cite{Witten1981,Guth1981,Yueker2014}) and can drastically increase the energy density in an expanding Universe. In the model in Ref.~\cite{Dvali:2016uhn}, this mechanism could give rise to relatively large neutrino masses even at a low apparent transition temperature, $T_{\Lambda_G}\lesssim \Lambda_G\sim v \sim m_\nu$. 

The relevant factors characterizing the possible delay of the phase transition are the mentioned neutrino asymmetries and unknown order-one coefficients in the effective potential $V(\Phi, T)$ of the neutrino-bilinear order parameters $\Phi\equiv\bar{\nu}\nu$. In the case of a strongly supercooled transition, the false metastable vacuum can be stabilized at $\langle \Phi\rangle=0$ over long cosmological times until tunnelling becomes significant at lower temperatures, which enables the false vacuum decay to the true minimum at $\langle \Phi\rangle\neq 0$ \cite{Mukhanov2005}. This vacuum decay releases positive potential energy density associated with the false vacuum and thus increases the energy density in the late relic neutrino sector relative to the other diluting energy densities in the Universe, e.g.\ of the photons. Consequently, the model in Ref.~\cite{Dvali:2016uhn} implies that the energy density in today's neutrino sector can be significantly larger than expected by standard cosmology.

Since a delayed neutrino phase transition would have a greater impact on cosmological observables than a non-delayed transition, the numerical analysis in this paper only focuses on the former case. In particular, Ref.~\cite{Koksbang:2017rux} found that in the case of neutrino mass generation at $T_{\Lambda_G}\sim \Lambda_G\sim v \sim m_\nu$, the cosmological limits are very similar to the constant-mass neutrino case if relic neutrino annihilation \cite{Dvali:2016uhn} is neglected. The neutrino masses will only slowly rise in this case, while the local minimum of the free energy will slowly decrease, with less impact on cosmological observations. In case of a supercooled phase transition, the neutrino masses and transition temperature are two independent parameters.

We note that generating relatively large masses at a low temperature seems to violate energy conservation at first sight. Therefore, differently from what was done in Ref.~\cite{Koksbang:2017rux}, we here take into account the false vacuum energy from the supercooled phase transition, which converts into neutrino masses at the low apparent transition temperature. Due to the unknown order-one coefficients in the effective potential mentioned above, the exact amount of false vacuum energy is an unpredictable free parameter of the theory. For simplicity, we assume that the false vacuum energy entirely converts into neutrino masses, and we neglect the additional conversion into excitations of the $\Phi$ field, i.e.\ dark radiation. 

We choose the same step-function parametrization for the late neutrino mass generation as Ref.~\cite{Koksbang:2017rux}
\begin{equation}
\label{eq:mnu}
m_{\nu}(a)=
\begin{cases}
0 
& 
\text{if $a \leq a_s$} \\
m_{\nu}\tanh{\Big(B_s\Big[\frac{a}{a_s}-1\Big]\Big)} 
&
\text{if $a>a_s$}
\end{cases}
\end{equation}
where $m_{\nu}$ is today's individual neutrino rest mass, $a$ is the scale factor, $a_s$ is the scale factor at the apparent phase transition time when the neutrino gains its mass, and $B_s$ is a parameter that determines the speed of the mass generation. We can fix the parameter $B_s$ to $10^{10}$, since the timescale of neutrino mass generation is of order $m_\nu^{-1}$, which corresponds to approximately femto/picoseconds.

We note that here we assume a degenerate neutrino mass spectrum, i.e.\ $m_{\nu_i}\equiv m_\nu$. Degenerate neutrino masses are still allowed in the mass model of Ref.~\cite{Dvali:2016uhn} because the standard cosmological mass limits are evaded, the bounds from $\beta$-decay experiments are relatively weak, and constraints from neutrinoless double-$\beta$ experiments only apply to Majorana neutrinos. Moreover, current cosmological data constrains only the sum of neutrino masses and cannot resolve yet whether the neutrino mass ordering is normal or inverted~\cite{Lattanzi:2017ubx,Jimenez:2010ev,Gerbino:2016ehw,Vagnozzi:2017ovm,Hannestad:2016fog,Capozzi:2017ipn,Gariazzo:2018pei}. Therefore, we assume degenerate masses that are generated at almost equal times for each mass eigenstate, i.e.\ within timescales much smaller than the Hubble timescale. Since the relic neutrinos rapidly decay into the lightest neutrino mass eigenstate, $\nu_l$, after the transition, the cosmologically constrained sum of the relic neutrino masses reduces to $\sum m_\nu= 3\times m_l$.

To model the time evolution of the false vacuum energy density, we can use a similar parametrization as for the neutrino mass above
\begin{equation}\label{eq:V0}
\rho_0(a)=
\begin{cases}
V_0\Big[1-\tanh{\Big(B_s\Big(1-\frac{a}{a_s}\Big)\Big)\Big]} 
& 
\text{if $a>a_s$} \\
V_0
&
\text{if $a \leq a_s$}
\end{cases}
\end{equation}
where $V_0=(\sqrt{p_\nu^2+m_\nu^2}-p_\nu)\,n_\nu$ is the difference in energy density of massive and massless neutrinos at $a=a_s$, and $n_\nu$ is the neutrino number density at that time. We assume here that the equation of state parameter of the false vacuum energy is constant, $w=-1$, and that only the amplitude of the energy density rapidly changes within timescales of femto/picoseconds, as discussed above. Therefore, the false vacuum energy effectively behaves as an additional vacuum energy contribution on top of dark energy, until the vacuum decays into the true minimum. Crucially, this scenario does not enhance the dark energy perturbations as in other mass-varying neutrino models~\cite{Franca:2009xp}, and hence is not affected by model instabilities. We notice here that we assume a standard cosmological constant for dark energy in our study and do not attempt to link it to the false vacuum energy. We will briefly comment on this in Sec.~\ref{sec:Discussions}.

\begin{figure}[t!]
	\includegraphics[width=\columnwidth]{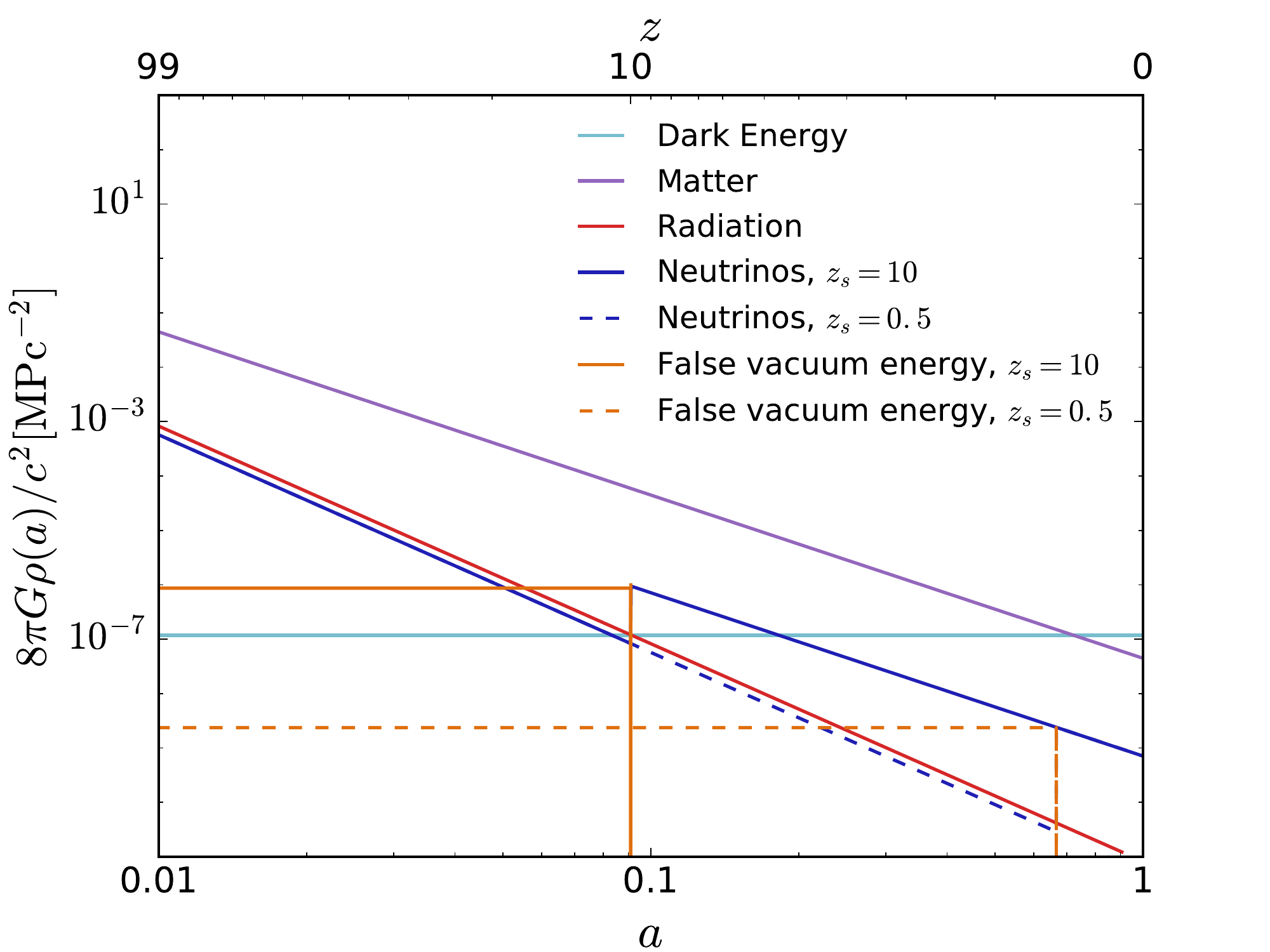}
	\caption{Energy densities for different components present in our analysis: neutrinos with a time-varying mass generated at $z_s=10$ ($z_s=0.5$) and corresponding to $\sum m_\nu=0.2$ eV today with solid (dashed) curves, false vacuum energy, standard dark energy, matter (baryons and cold dark matter), and radiation.}
	\label{fig:energy}
\end{figure}

The energy densities of massive neutrinos and the false vacuum energy component are shown in Fig.~\ref{fig:energy} for $\sum m_\nu=0.2$ eV and a late phase transition at a redshift of $z_s=10$ (or equivalently $a_s\sim0.091$, solid lines). In this case, the false vacuum energy dominates over the dark energy density until the phase transition and is then transferred into the energy required for the generation of the neutrino masses. We also show with dashed lines the case of a very late phase transition happening at $z_s = 0.5$, we note that in this case the false vacuum energy is more subtle and always subdominant compared to dark energy.

\subsection{Cosmological Observables}
\begin{figure*}[ht!]
	\includegraphics[width=0.95\columnwidth]{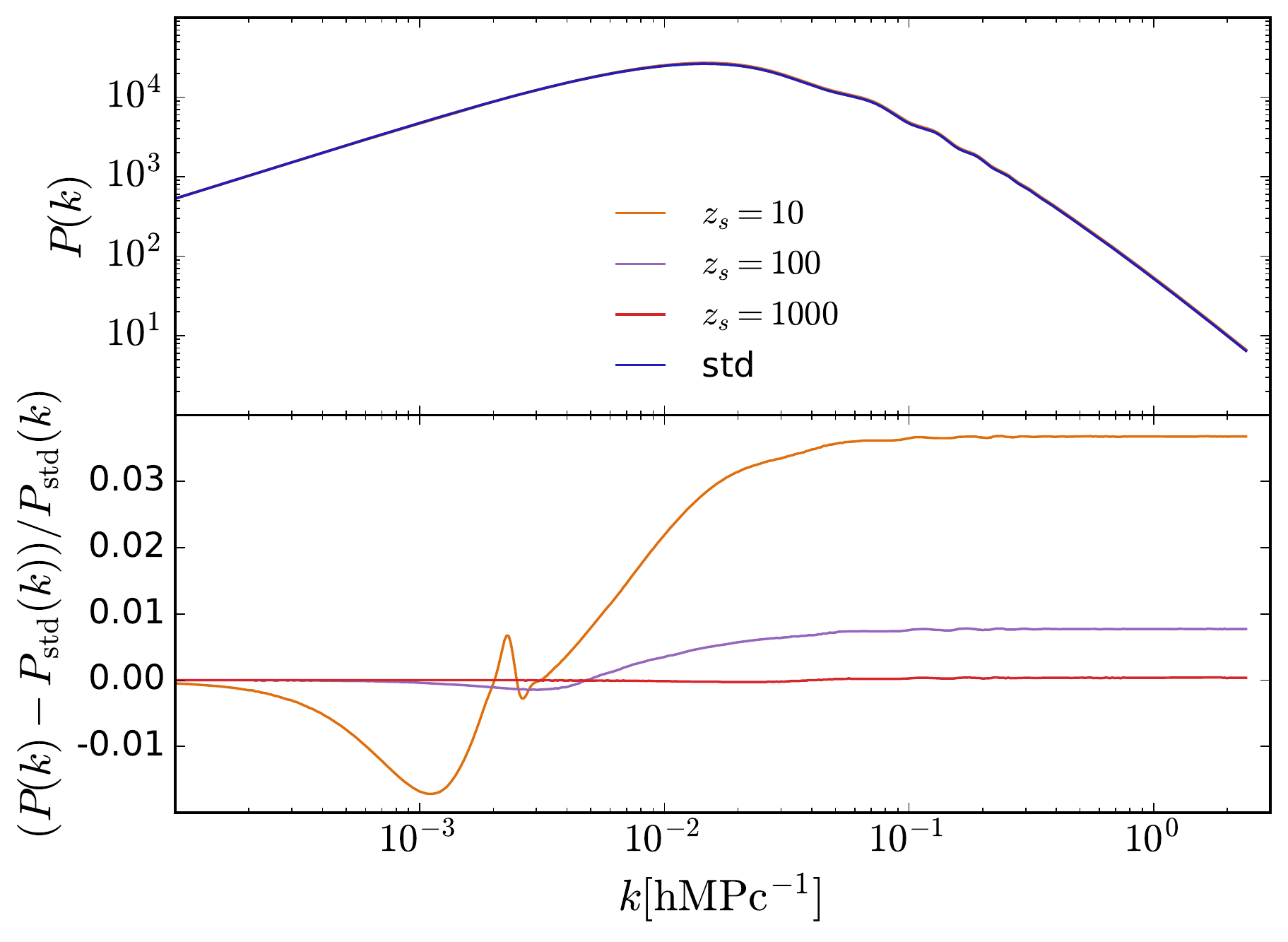}
	\includegraphics[width=0.97\columnwidth]{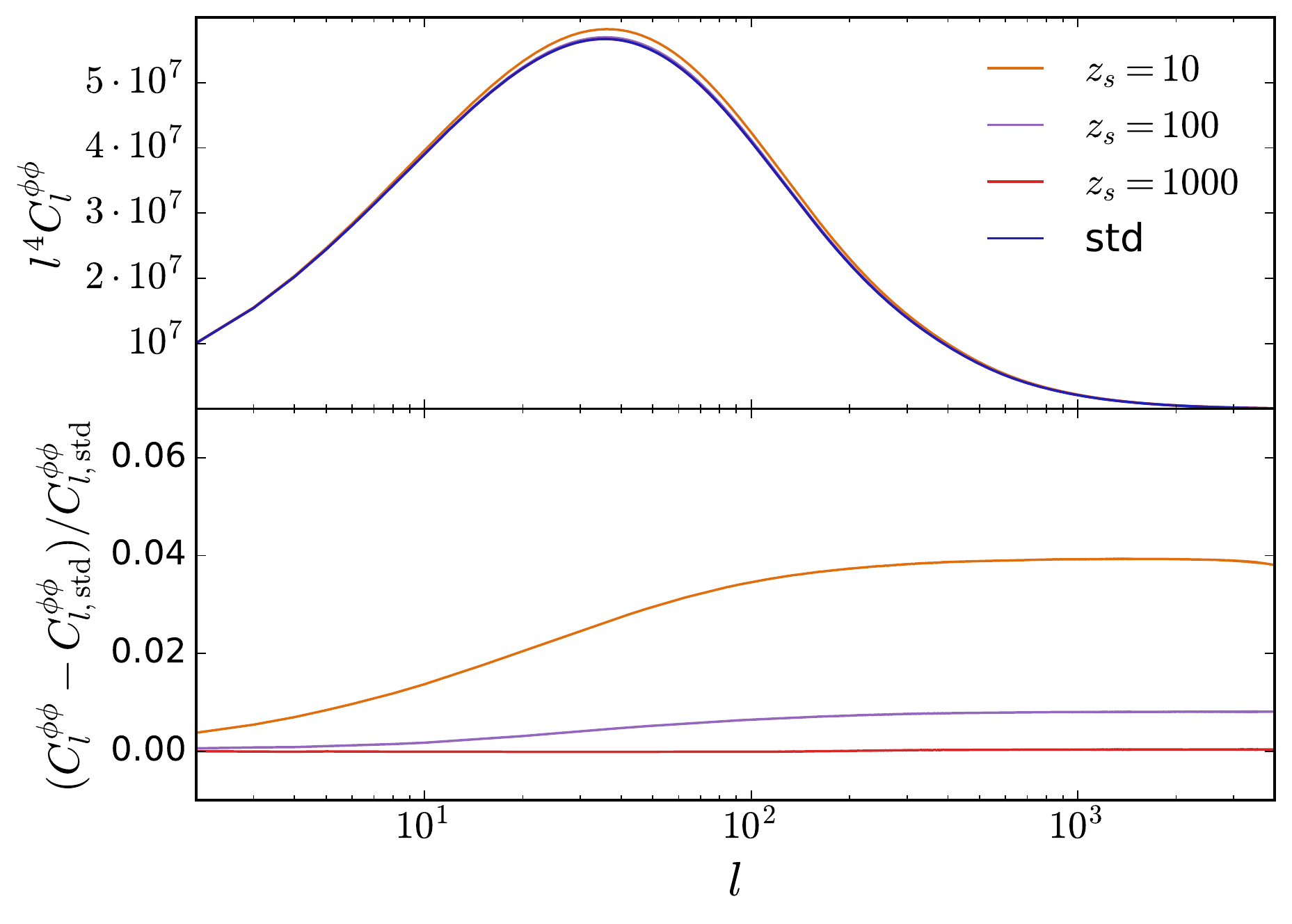}
	\includegraphics[width=0.95\columnwidth]{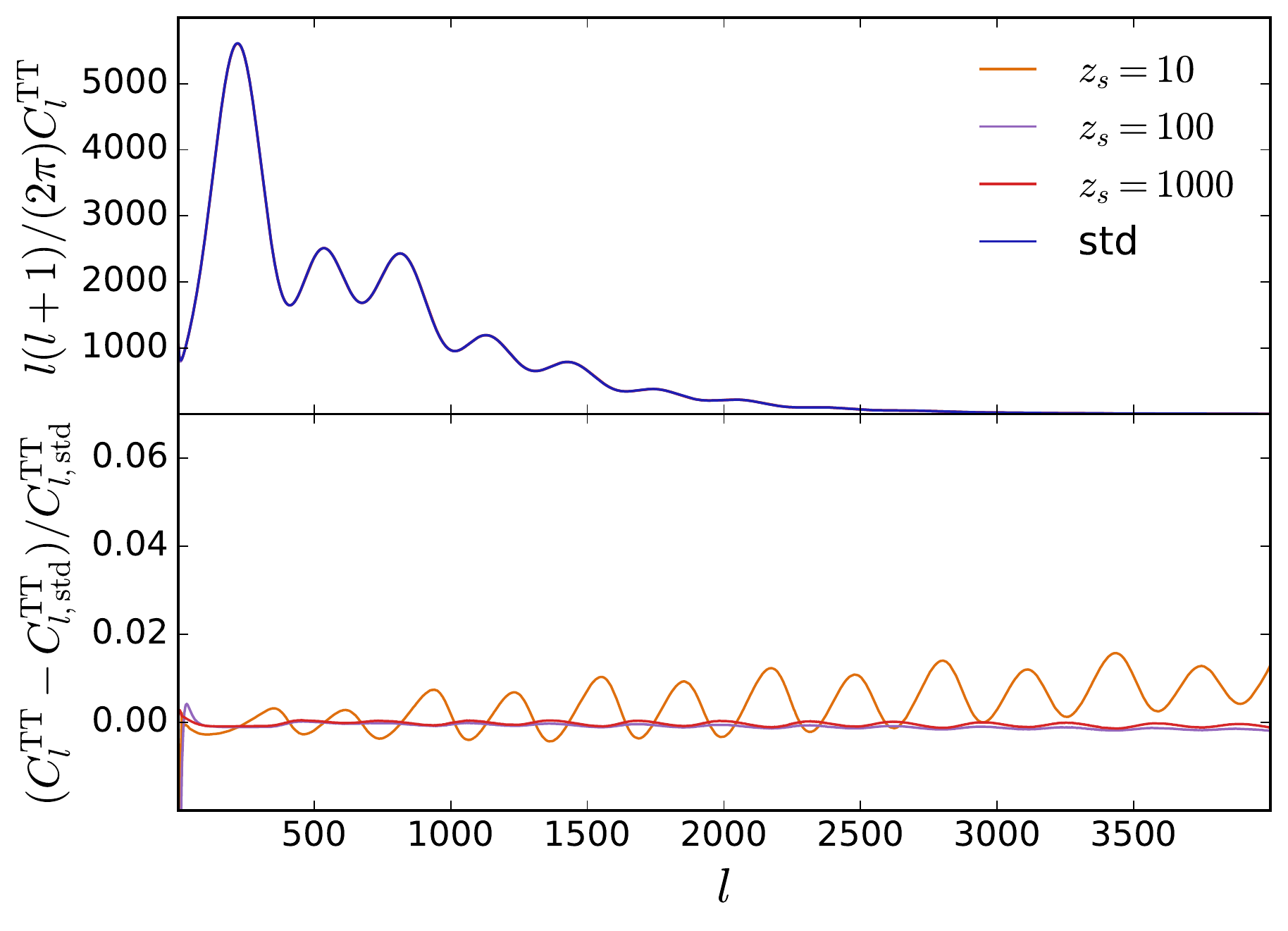}
	\includegraphics[width=0.96\columnwidth]{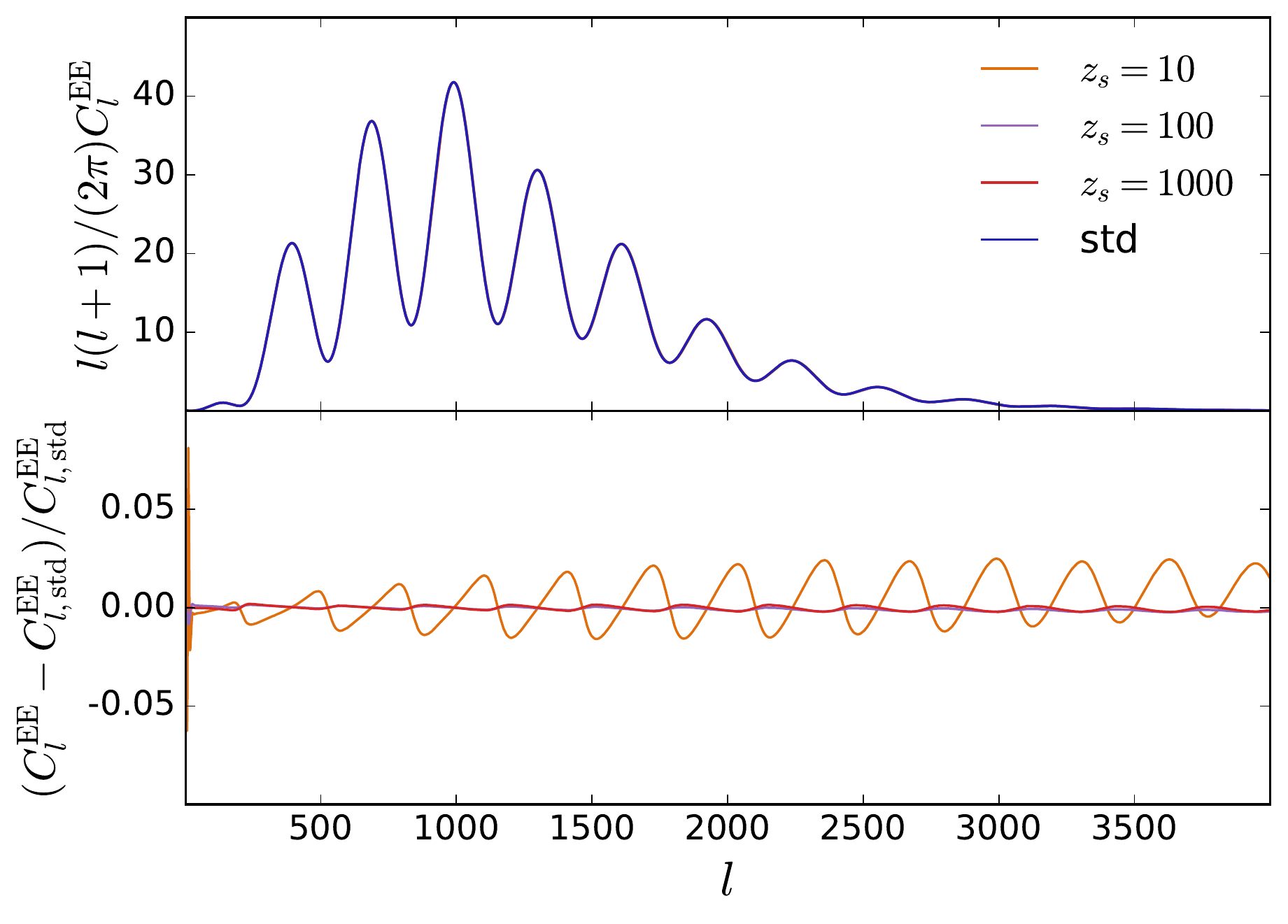}
	\caption{Effect on cosmological observables from the time-varying neutrino mass model considered here, shown for $\sum m_\nu=0.2$~eV and for three different values of the phase transition redshift ($z_s=1000$ or $a_s=0.001$, $z_s=100$ or $a_s=0.01$ and $z_s=10$ or $a_s$=0.09), compared to the standard massive neutrinos case with $\sum m_\nu=0.2$~eV used as a reference. Different panels report the matter power spectrum (top left), the CMB lensing convergence power spectrum (top right), and CMB temperature and polarization anisotropy power spectra (bottom left and right, respectively). The effects of this model are subtle, with percent level features, but within the reach of future experiments.}
\label{fig:cmb}
\end{figure*}

\begin{figure*}[ht!]
	\includegraphics[width=0.95\columnwidth]{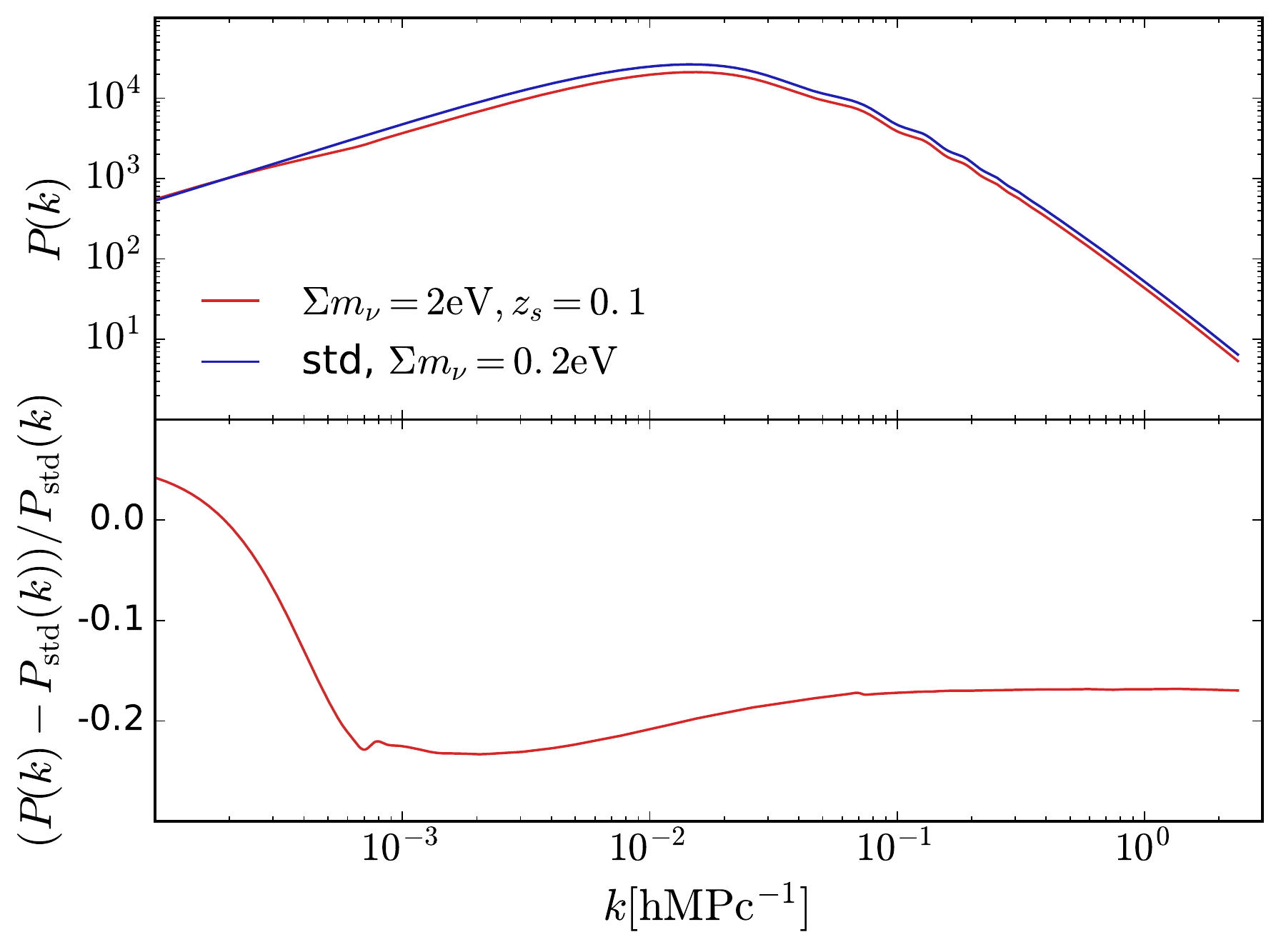}
	\includegraphics[width=0.99\columnwidth]{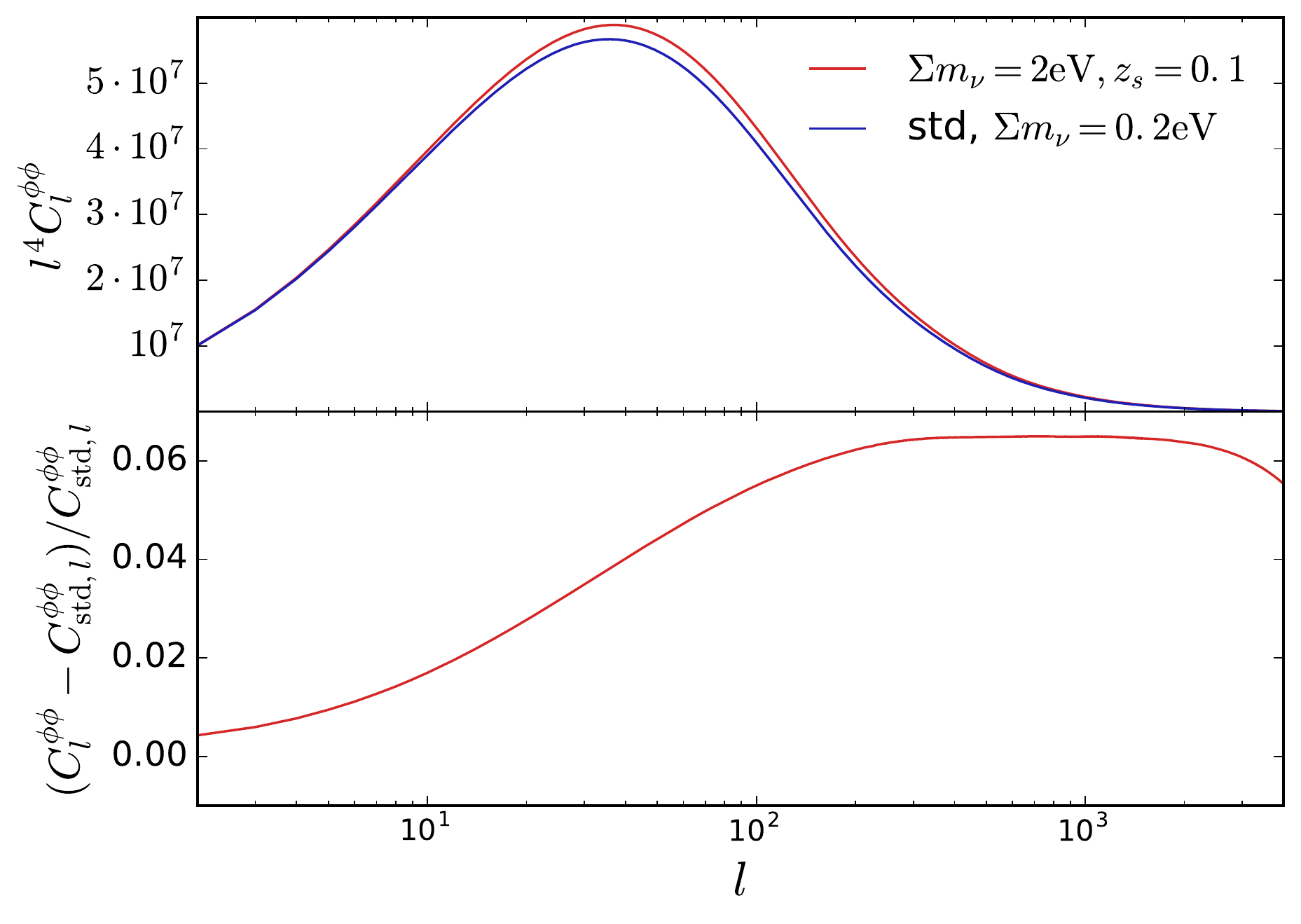}
	\includegraphics[width=0.95\columnwidth]{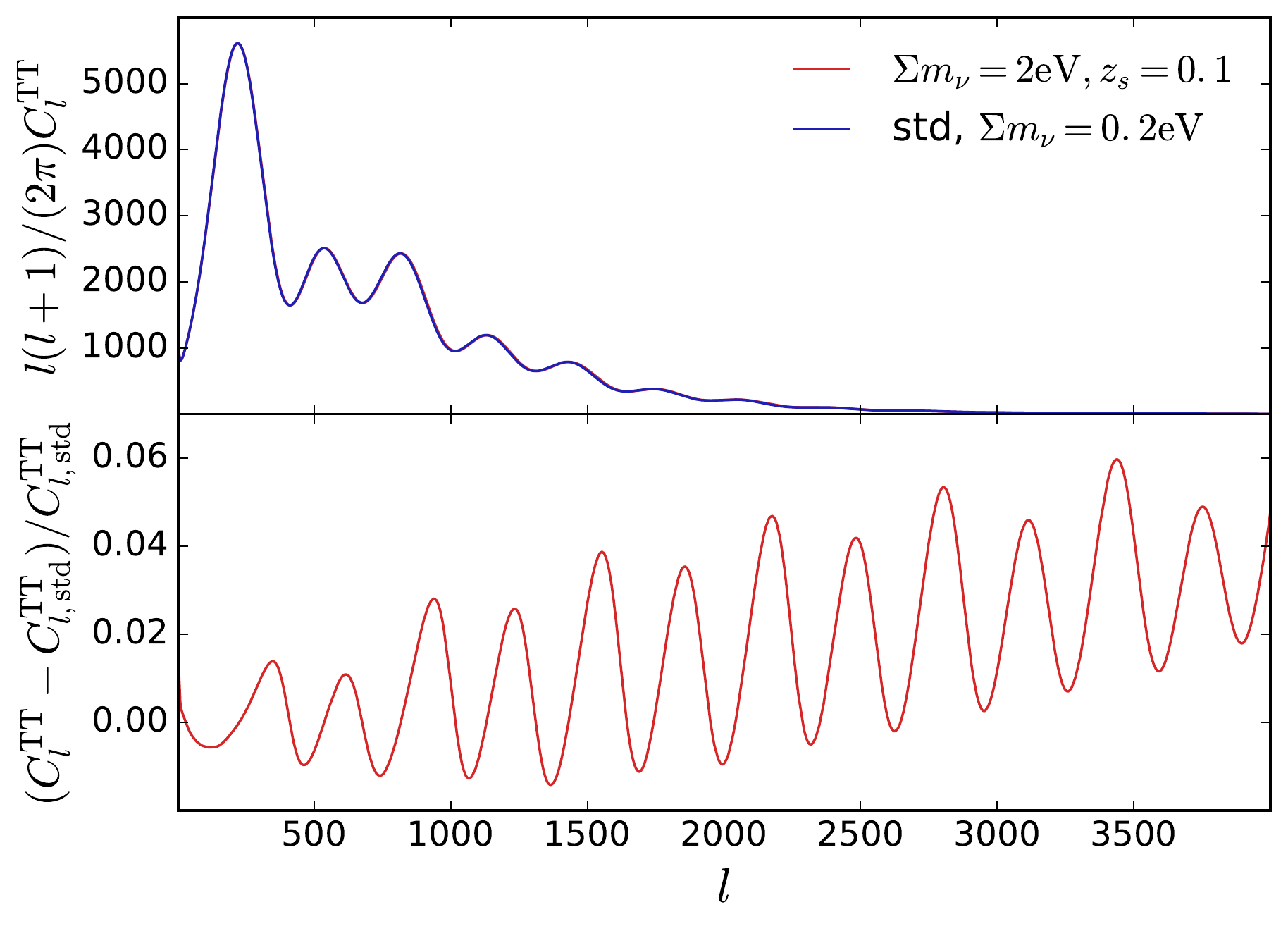}
		\includegraphics[width=0.96\columnwidth]{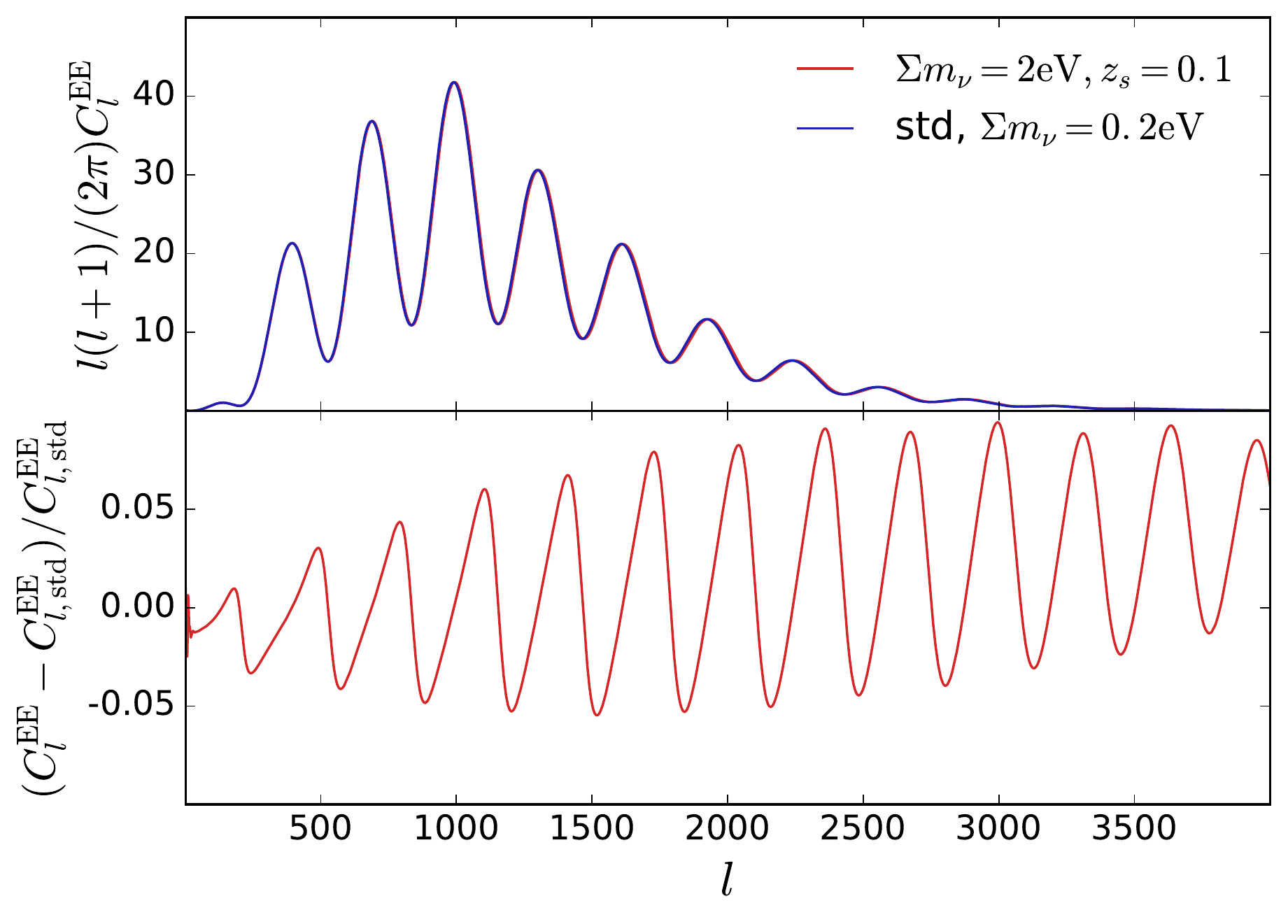}
	\caption{Same as Fig.~\ref{fig:cmb} in the case of a late phase transition ($z_s=0.1$ or $a_s=0.91$) and a large neutrino mass with $\sum m_\nu=2$ eV, compared to the standard massive neutrinos case with $\sum m_\nu=0.2$~eV.}
	\label{fig:cmb_high_nu_mass}
\end{figure*}

The impact of this model on cosmological observables is shown in Figs.~\ref{fig:cmb},~\ref{fig:cmb_high_nu_mass}: features in the CMB temperature ($C_l^{\rm TT}$) and polarization ($C_l^{\rm EE}$) power spectra, the CMB lensing convergence ($C_l^{\phi\phi}$) and matter power spectra ($P(k)$) are shown for $\sum m_\nu=0.2$ eV and three different values of $z_s$ (Fig.~\ref{fig:cmb}), and for a late phase transition and high neutrino masses (Fig.~\ref{fig:cmb_high_nu_mass}), with the standard massive neutrino case as the reference model in both cases. 

When the phase transition happens late (small values of $z_s$ and large values of $a_s$), the model becomes more similar to massless neutrinos. On the contrary, for large values of $z_s$, the model is very similar to the standard constant mass neutrino case. Therefore, for $z_s=1000$ ($a_s=0.001$), the effect of the time-varying neutrino mass for all four power spectra is only marginal compared to the reference case. 

We start our explanation with the matter power spectrum on the top left corner of Fig.~\ref{fig:cmb}. 
For standard massive neutrinos, the matter power spectrum is suppressed on small scales, in the case of $\sum m_\nu=0.2$ eV this corresponds to $k\geq k_\mathrm{nr}=0.0027$. This suppression is more or less pronounced in our case depending on the time of the phase transition. As mentioned above, for a small value of $z_s$ the neutrinos are massless for most of their evolution and as a result the matter power spectrum is less suppressed and more similar to the power spectrum of massless neutrinos. As described in Ref.~\cite{Koksbang:2017rux}, the turn-over-scale of the matter power spectrum is also affected, depending on the exact time at which the neutrinos gain their mass. As the time of the phase transition moves towards smaller redshifts, the enhancement of the matter power spectrum for large values of $k$ translates into an overall enhancement of the lensing convergence power spectrum (top right panel of Fig.~\ref{fig:cmb}). 
The CMB temperature and polarization anisotropy power spectra (bottom left and right panel, respectively) are mostly affected at small and large multipoles, encoding the impact of extra vacuum energy and neutrino free-streaming in the case of a late phase transition. 

Anticipating larger values of $\sum m_\nu$ allowed by a supercooled phase transition, in Fig.~\ref{fig:cmb_high_nu_mass} we compare cosmological observables in the case of mass-varying neutrinos with $\sum m_\nu=2$~eV with respect to the standard massive neutrino case with $\sum m_\nu=0.2$~eV. We notice that even in the case of these very different mass scenarios the impact on the observables is subtle. For this comparison, we have not renormalized the values of the different matter density components (i.e.\ we kept the amount of cold dark matter and baryons fixed) to reproduce the process where neutrinos exchange some energy only with the false vacuum energy component (i.e.\ moving along the dark energy degeneracy line seen in Sec.~\ref{sec:Limits}). A higher impact is now seen on $P(k)$, showing a suppression on all scales out to the horizon at the phase transition scale, caused by the substantial amount of false vacuum energy before the transition. The features in the CMB spectra are also enhanced due to the different energy budget of the Universe.

We note that the differences between the models are only of the order of a few percent. We anticipate that this might be hard to uncover with current data but is within the reach of future CMB and galaxy surveys. The CMB SO~\cite{SO} and Stage-4 projects~\cite{CMB-S4} will have the sensitivity to distinguish the small-scale CMB features,  while Euclid~\cite{Euclid} and LSST~\cite{LSST} will provide better measurements of $P(k)$.
 
\section{Analysis methodology}\label{sec:analysis}
To constrain the parameters of our model, we use modified versions of the publicly available Boltzmann solver CAMB~\cite{Lewis:1999bs} and the Monte-Carlo Markov chain package CosmoMC~\cite{Lewis:2002ah}. We compare this model where neutrino masses are generated through a supercooled phase transition, named hereafter {\tt Supercool}-$\nu$, to the standard $\Lambda$CDM case with fixed neutrino masses $\sum m_\nu=0.06$~eV, and to the case in which the total mass is varied but constant in time (i.e.\ the standard massive neutrino case), $\Lambda$CDM+$\sum m_{\nu}$. 

When reporting $\Lambda$CDM results, we vary the standard six cosmological parameters (the baryon and cold dark matter densities, $\Omega_b$ and $\Omega_c$, the scalar spectral index $n_s$, the amplitude of primordial fluctuations, $A_s$, the Hubble constant, $H_0$, and the optical depth to reionization, $\tau$) and fix the total sum of neutrino masses to $\sum m_\nu=0.06$ eV, corresponding approximately to the lower limit obtained from neutrino oscillation experiments~\cite{Tanabashi}. In the extended analyses for i) $\Lambda$CDM+$\sum m_{\nu}$ we additionally vary $\sum m_\nu$ as a constant parameter; and for 
ii) the {\tt Supercool}-$\nu$ model we additionally consider the full time evolution of the neutrino mass and vary the scale factor of the phase transition, $a_s$. The false vacuum energy amplitude is set by the value of $\sum m_{\nu}$ and $a_s$ via Eq.\ \eqref{eq:V0}. 

Unless otherwise stated (for example in Section~\ref{sec:Tensions}), we assume standard flat priors on the $\Lambda$CDM basic parameters (following Ref.~\cite{planck2015xiii}). We vary $\sum m_\nu$ between 0.06 and 6.6 eV to incorporate current limits from laboratory searches (i.e.\ above the minimum threshold set by oscillation experiments and converting $m_{\nu_e}<2.2$~eV into $\sum m_\nu<6.6$~eV). We will extend this range in Sec.~\ref{sec:Tensions} to ease the comparison with other published results.
The logarithm of the time of the phase transition, $\log (a_s)$, is varied between -5 and 0. This allows the exploration of neutrino mass generation across a large range of cosmic time. We fix the speed of the transition with $B_s=10^{10}$, corresponding to an almost instantaneous phase transition. This parameter was very unconstrained in the analysis of Ref.~\cite{Koksbang:2017rux}, so we do not expect its exact value to affect our results. 

We separate our analysis in two parts: in Sec.~\ref{sec:Limits} we report state-of-the-art constraints for the parameters of the time-varying neutrino mass model considered here; in Sec.~\ref{sec:Tensions} we study the constraints in the $\Omega_m$-$\sigma_8$ plane from different cosmological probes. 
 
\section{Results}
\subsection{Cosmological Mass Limits}\label{sec:Limits}

\begin{figure}[t!]
	\includegraphics[width=\columnwidth]{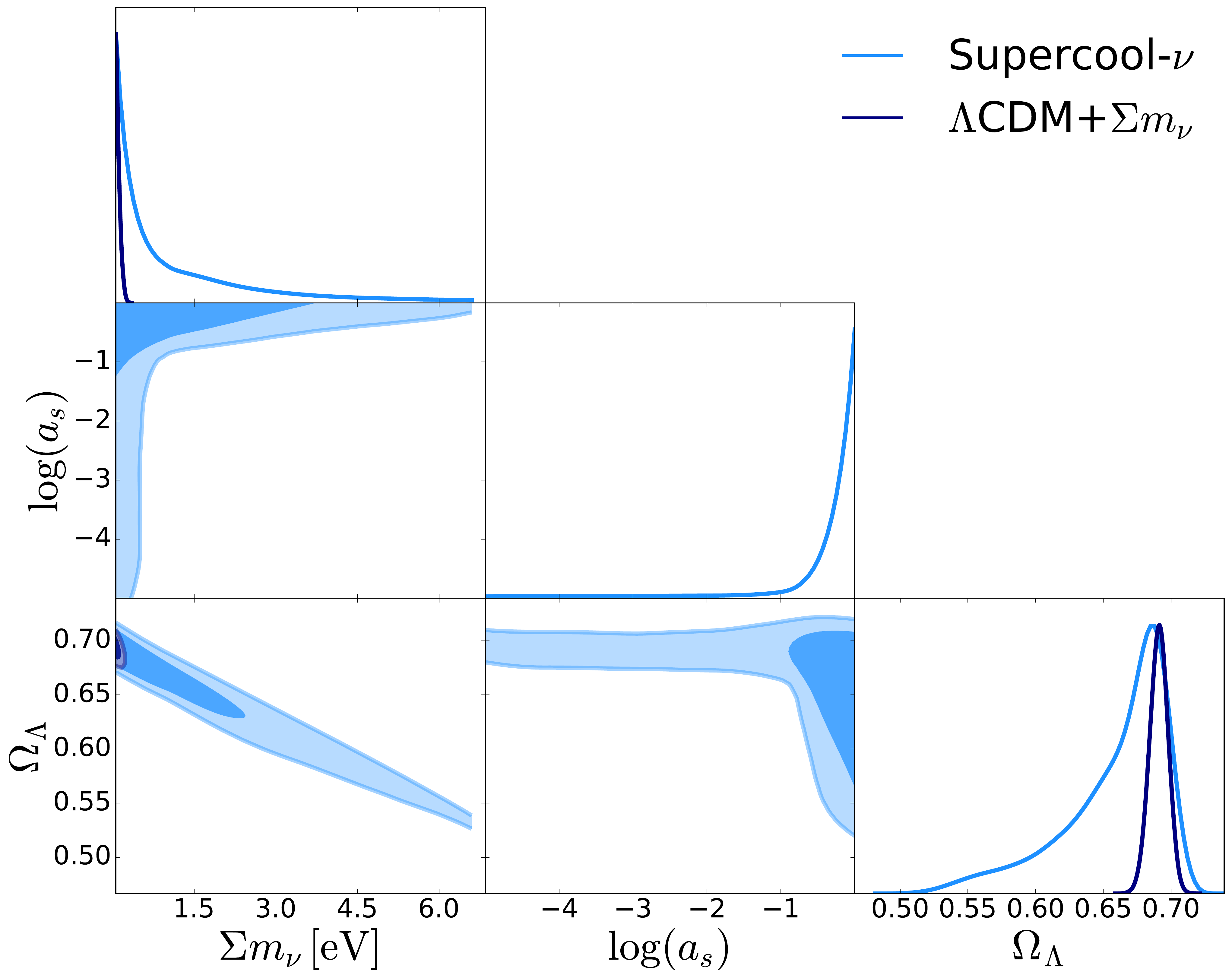}
	\caption{Constraints for $\sum m_\nu$, $a_s$ and $\Omega_\Lambda$ (with contours at 68\% and 95\% confidence) in the case of standard massive neutrino ($\Lambda$CDM+$\sum m_{\nu}$, dark blue) or for relic neutrinos with mass generated in a supercooled phase transition ({\tt Supercool}-$\nu$, light blue). The results are obtained using \textit{Planck} TTTEEE+lensing, BAO and SN.} 
	\label{fig:mnu_limits}
\end{figure}

To obtain constraints from current data, we combine \textit{Planck} CMB temperature, polarization, and lensing spectra from the 2015 release~\cite{Aghanim:2015xee,Ade:2015zua}\footnote{The final 2018 \textit{Planck} release occurred during the final stages of this work. We, however, note that the 2018 likelihood software needed to analyse the data is not yet public, and we anticipate that our results will not change with the new data products.} with BAO distance ratio from BOSS DR12 (CMASS and LOWZ)~\cite{2015ApJS..219...12A}, SDSS MGS~\cite{Ross:2014qpa} and 6DF~\cite{2011MNRAS.416.3017B}, and Type Ia supernovae redshift-magnitude diagram from the Joint  Light-curve  analysis  (JLA) compilation~\cite{jla}. This is the baseline data combination of the \textit{Planck} analyses that we follow here. The results are shown in Fig.~\ref{fig:mnu_limits} and reported in Table~\ref{tab:limits}.
\begin{table}[t]
\begin{tabular}{l|c|c}
Parameters & $\Lambda$CDM+$\sum m_\nu$ & \tt{Supercool-}$\nu$ \\
\hline
\hline 
$\sum m_\nu$[eV] & $\leq 0.20$ & $\leq 4.8$ ($\leq 1.6$)\\
$\Omega_\Lambda$ & $0.69\pm0.01$ & $0.66_{-0.04}^{+0.02}$\\
$\log(a_s)$ & --- &  $\geq -3.6$ ($\geq -2.8$) \\
\hline
\hline
\end{tabular} 
\caption{\label{tab:limits} Marginalized constraints on the sum of neutrino masses and dark energy content today, and on the scale factor of the neutrino mass generation using \textit{Planck} CMB temperature, polarization and lensing, BAO  and SN data. Errors are given at 68\% confidence, and upper/lower limits are reported at 95\% confidence (and also at 68\% confidence in parentheses for very non-Gaussian bounds).}
\end{table}

We find that much larger values for $\sum m_\nu$ are allowed in the case of a supercooled phase transition compared to the case of standard constant-mass neutrinos and that the data prefer a large value of the phase transition scale factor, i.e.\ a late relic neutrino mass generation (peaking at today's scale factor)
\begin{equation}
\begin{rcases}
\sum m_\nu & \leq  4.8~{\rm eV}\\
\rm{log}(a_s) & \geq  -3.6 
\end{rcases}~{\rm at~ 95\% ~confidence}\,.
\end{equation}

This is a significantly weakened limit for the neutrino mass, to be compared to $\sum m_\nu\leq0.2$~eV for standard massive neutrinos with the same data combination -- we note though that the 68\% limit, $\sum m_\nu\leq1.6$~eV, is much tighter due to the non-Gaussian distribution recovered in this fit. This is expected in this model and the reason for this is illustrated in Fig.~\ref{fig:energy}: the inclusion of the false vacuum energy generates a condition where the amplitude of the dark energy density and the combination of the neutrino and false vacuum energy components are very similar over most of the cosmic history. Especially in the case when the transition happens very late ($z_s \leq 10$) and the sum of neutrino masses is large, the neutrino energy density will be of the same order of magnitude as the dark energy density until almost today. Therefore, a strong anti-correlation between $\sum m_\nu $ and $\Omega_\Lambda$ arises (at the level of 98\%). This can also be seen in Fig.~\ref{fig:mnu_limits}. A similar degeneracy has also been observed for early dark energy (EDE) models~\cite{Lorenz:2017fgo,2011PhRvD..83l3504C}, however in these models the degeneracy is caused by the time-varying evolution of the dark energy component. We also note that the correction that we added to keep energy conserved in the model, i.e.\ the inclusion of the false vacuum energy, is the main reason why our constraints are broader than those reported in Ref.~\cite{Koksbang:2017rux}. 
The preference for a late transition captures the trend that has emerged fitting for neutrino masses with early- and late-time cosmological probes: we confirm that the data require lighter neutrinos at CMB decoupling and more significant masses can be generated only in the late Universe. 

The goodness of the fit obtained with this time-varying neutrino mass model is only marginally better than that obtained in the standard massive neutrino case, with a difference in best-fit likelihoods of only 1.57 ($\Delta\chi^2=3.14$). Therefore, the {\tt{Supercool-}$\nu$} model is slightly but not significantly favoured, yielding a \textit{p}-value of 0.08 with one additional degree of freedom for the {\tt{Supercool-}$\nu$} model compared to $\Lambda$CDM$+\sum m_\nu$.

\subsection{The $\boldsymbol{\Omega_m}$-$\boldsymbol{\sigma_8}$ plane}\label{sec:Tensions}

We now compare cosmological constraints in the $\Omega_m$-$\sigma_8$ plane. We use \textit{Planck} CMB temperature and polarization data, \textit{Planck} lensing and \textit{Planck} SZ cluster counts data~\cite{Ade:2015fva}, and galaxy weak lensing data from KiDS~\cite{Hildebrandt:2016iqg}\footnote{We work with KiDS weak lensing data because this is the most discrepant data and because it was the only publicly available likelihood at the time this work started.}. We take each dataset singularly, except for the SZ case where, following the \textit{Planck} analysis, we further add BBN constraints on $\Omega_bh^2$ to break parameter degeneracies. This choice is made to explore the impact of time-varying neutrino masses on the existing tensions, and whether the inclusion of massive neutrinos generated late in the Universe might ease the discrepancies.  \textit{Planck} CMB lensing is also included in our analysis as a dataset on its own, not because of tension with other data but rather to look at the effect of this model at intermediate-to-low redshifts. 

To easily compare with the KiDS weak lensing, \textit{Planck} lensing, and \textit{Planck} SZ cluster results, we use now the same flat priors for the unconstrained parameters assumed by the individual experiments. For KiDS we use the priors assumed in Ref.~\cite{Joudaki:2016kym}; for \textit{Planck} CMB lensing we use the priors for $n_s$ and $\Omega_b$ as stated in Ref.~\cite{Ade:2015zua}; and for the \textit{Planck} SZ cluster counts we use the priors for $n_s$ and $\Omega_b$ reported in Ref.~\cite{Ade:2015fva}. For these latter data we further assume a Gaussian prior on the bias parameter picking the CCCP baseline case~\cite{Hoekstra:2015gda} used as reference cluster mass calibration in the \textit{Planck} analyses. For the galaxy weak lensing, \textit{Planck} lensing and SZ cases $\tau$ is not varying. We note that the neutrino mass parameter is now varied between 0.06 and 10~eV consistently with other published analyses and therefore for a simpler comparison.

\begin{figure}[t!]
	\includegraphics[width=\columnwidth]{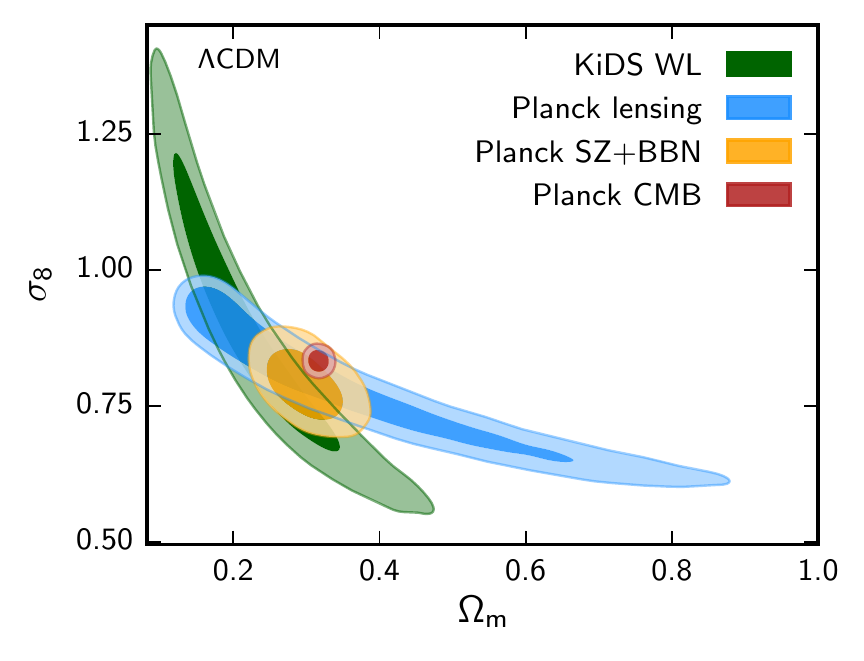}
	\includegraphics[width=\columnwidth]{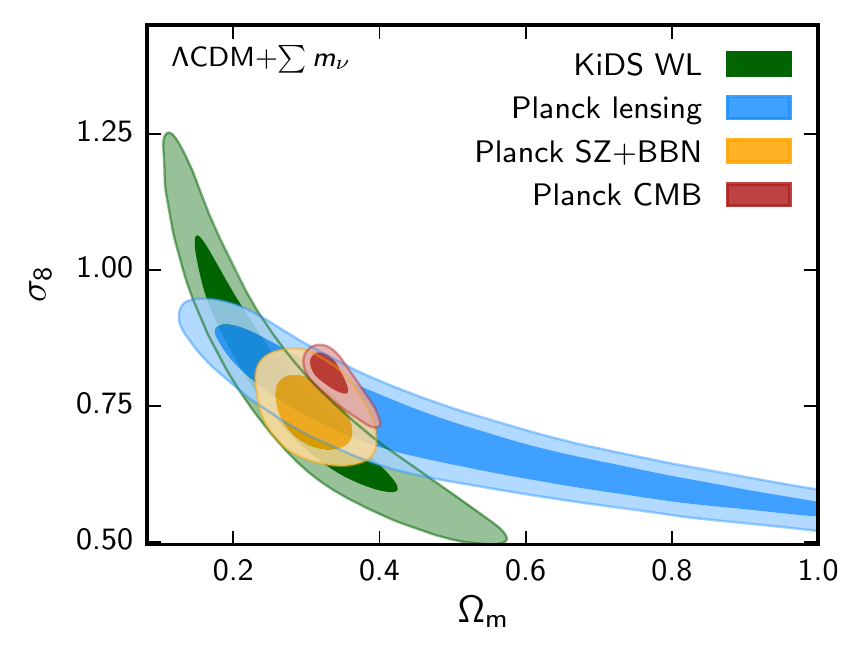}
	\includegraphics[width=\columnwidth]{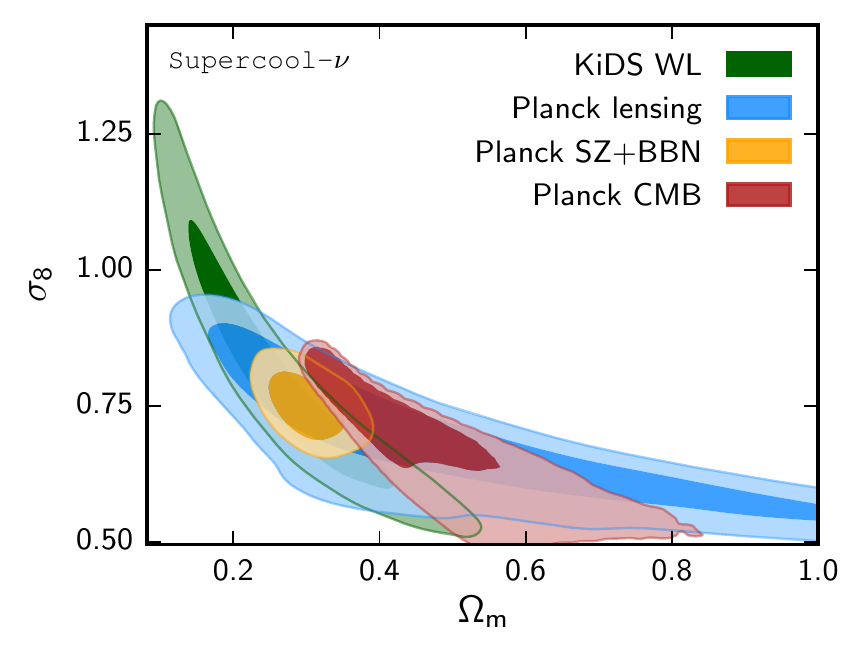}
	\caption{Constraints on $\sigma_8$ and $\Omega_m$ inferred from \textit{Planck} CMB, \textit{Planck} lensing, \textit{Planck} SZ cluster counts and KiDS weak lensing. \emph{Top:} for $\Lambda$CDM with fixed neutrino mass $\sum m_\nu=0.06$~eV. \emph{Middle:} $\Lambda$CDM with neutrino mass $\sum m_\nu$ as a constant free parameter. \emph{Bottom:} late neutrino mass generation with $\sum m_\nu$ and $a_s$ as a free parameter.
}		
	\label{fig:tensions}
\end{figure}

The results for the three models compared here are shown in Fig.~\ref{fig:tensions} (transition from top to bottom) and discussed below.\\

\noindent {\bf $\Lambda$CDM:} The top panel of Fig.~\ref{fig:tensions} shows the constraints for $\Omega_m$ and $\sigma_8$ in the case of $\sum m_\nu$ fixed to $0.06$ eV for the four different datasets considered here. These results reproduce the published KiDS~\cite{Joudaki:2016kym}, {\textit{Planck} CMB and CMB lensing~\cite{planck2015xiii,Ade:2015zua}, and \textit{Planck} SZ+BBN~\cite{Ade:2015fva}\footnote{We have cross-checked our \textit{Planck} SZ+BBN results by additionally including BAO and comparing with the \textit{Planck} SZ+BBN+BAO constraints in Ref.~\cite{Ade:2015fva}.} results, and are shown here only for reference\footnote{We note that these contours will shift if using different $\tau$ values compared to the \textit{Planck} 2015 one used here. However, we do not expect this to change significantly any conclusion drawn in this paper. We decided to keep the 2015 value to compare more easily with other published results.}.\\

\noindent {\bf $\Lambda$CDM+$\sum m_\nu$:} When varying $\sum m_\nu$ as a parameter, correlations in the matter components generate a broadening of the constraints. In particular, the middle panel of Fig.~\ref{fig:tensions} shows the impact of the standard massive-neutrinos-driven suppression of density fluctuations below their free-streaming length. Larger allowed values for the neutrino mass enlarge the \textit{Planck} CMB primary and lensing constraints towards lower values of $\sigma_8$ and higher values of $\Omega_m$. Similar effects are seen for the KiDS and SZ analysis. This has been extensively demonstrated in the literature (e.g.\ \cite{Joudaki:2016kym,Battye:2013xqa,Wyman:2013lza,Salvati:2017rsn}).\\

\noindent{\tt Supercool}-$\nu$: The bottom panel in Fig.~\ref{fig:tensions} shows our results for the supercooled phase transition. The largest impact compared to the other two cases is seen on the \textit{Planck} CMB contours: they now extend to much lower values of $\sigma_8$ and higher values of $\Omega_m$. CMB lensing contours are slightly affected, while the KiDS and SZ cluster results are almost unchanged. This is explained by the data preferring a late-time mass generation, so that the {\tt Supercool}-$\nu$ case only differs significantly from the $\Lambda$CDM or $\Lambda$CDM+$\sum m_\nu$ cases at CMB and CMB lensing epochs. The contours however broaden along the degeneracy line, bringing data in slightly better agreement but with no substantial model preference (when considering the broadening due to the extra parameters present in the model). We also note that the derived value of the Hubble constant in this model is not significantly different from the one obtained in the $\Lambda$CDM+$\sum m_\nu$ case.}

\section{Summary and concluding remarks}\label{sec:Discussions}

In this paper, we have presented state-of-the-art constraints from cosmology on a time-varying neutrino mass model motivated by Ref.~\cite{Dvali:2016uhn}. We assume that relic neutrino masses are generated from a form of false vacuum energy in a supercooled neutrino phase transition and neglect neutrino annihilation in the late Universe. This is a modified version of the minimal model in Ref.~\cite{Dvali:2016uhn} which predicts almost complete neutrino annihilation after the supercooled phase transition and thus implies that all cosmological mass constraints are entirely evaded. We find that current data prefer a phase transition very late in time (peaking at today) and that the constraint on the total mass of neutrinos is significantly weakened compared to the standard massive neutrinos case, with $\sum m_\nu \leq 4.8$~eV at 95\% confidence ($ \leq 1.6$~eV at 68\% confidence). This larger bound is mostly due to large correlations with the dark energy component, affected by the presence of the false vacuum energy term. To summarize, we find that the standard constant-mass neutrino case with low masses and the {\tt Supercool}-$\nu$ model studied here with high masses are both successful with current data.
 
The recently proposed PTOLEMY experiment \cite{Betts:2013uya} aims to achieve the sensitivity required to detect relic neutrinos. However, such a detection would only be feasible in case of degenerate or quasi-degenerate neutrino masses due to the proposed energy resolution of $\sim 0.15$~eV per neutrino~\cite{Long:2014zva}. While such large masses are ruled out by conventional cosmological neutrino mass bounds, the results found here still allow for a detection by PTOLEMY in the presence of a strongly asymmetric neutrino background, as we will further discuss below. The KATRIN $\beta$-decay experiment \cite{Drexlin:2013lha} also has the potential to discover a relatively large absolute neutrino mass scale soon. Since the model considered here allows for larger neutrino masses, a detection of an unexpectedly large absolute neutrino mass scale at KATRIN could provide a strong hint towards this model, at least if the standard cosmological $\Lambda$CDM model is valid in other respects. We note that the KATRIN measurement would not be affected by possible modifications of the measured electron energy spectrum due to neutrino self-interactions, since the $\beta$-decay process happens on much shorter timescales than these interactions. The weakened neutrino mass bounds gain even further importance in the hypothetical presence of sterile neutrinos motivated by experimental short-baseline anomalies \cite{Gariazzo2015}. Light sterile neutrinos usually stand in conflict with cosmological bounds on neutrino masses and the primordial radiation density \cite{Adhikari2016}, but these conflicts vanish in the model~\cite{Dvali:2016uhn}, since the relic (active) neutrinos are massless in the early Universe and thus have vanishing couplings to their sterile partners.

We further looked at the possibility of solving current early- and late-time tensions in the measurements of matter fluctuations with this model. Larger values allowed for the neutrino mass also weaken constraints on the matter density and clustering. These, however, broaden along the degeneracy direction already present in the standard constant mass case and do not provide a convincing explanation to the tensions. \\
 
We made several simplifications to the original neutrino mass model in Ref.~\cite{Dvali:2016uhn}. 
\begin{itemize}
\item The model predicts that the relic neutrinos rapidly decay into the lightest neutrino mass eigenstate after the late cosmic transition. Therefore, any cosmological neutrino mass bound derived with this model only applies to the smallest neutrino mass and not to the sum of all masses. Considering that, at present, we do not have further information on the neutrino mass eigenstates ordering and relative weight, we argue that making this simplification is not impacting our conclusion. Moreover, the decay becomes less relevant for larger masses, since then the neutrino mass eigenstates have similar masses and are cosmologically not distinguishable. We also note that the relic neutrino decay into the lightest mass eigenstate results in an enhanced (suppressed) relic neutrino detection rate at PTOLEMY for a normal (inverted) neutrino mass ordering, because the lightest mass eigenstate contains a large (small) fraction of the electron neutrino flavor eigenstate. 

\item The model in Ref.~\cite{Dvali:2016uhn} also predicts that the relic neutrinos become strongly coupled after the phase transition and substantially annihilate into almost massless Goldstone bosons, i.e.\ dark radiation. In the case of almost complete annihilation, this would not be tracked by neutrino masses from cosmological data. We relax this prediction by Ref.~\cite{Dvali:2016uhn} for two reasons: i) first, an evidence of time-varying neutrino masses from cosmology could still inform model building in general. Our study confirms the general trend that low-redshift data prefer heavier neutrinos and showed that large masses can be generated only in the late Universe. We note here that the latter result is expected to also hold true in case of complete neutrino annihilation, due to the larger amount of false vacuum energy required for an earlier phase transition. ii) An almost complete annihilation could in fact be evaded in the presence of large neutrino asymmetries and could be falsified by a cosmological neutrino mass detection. We showed that non-complete annihilation is still a viable possibility considering the current bounds on these asymmetries.

\item Another aspect we neglected in our study is the formation and evolution of topological defects, as well as out-of-equilibrium effects like bubble nucleation and collision. Related cosmological studies of the resulting inhomogeneities in supercooled late-time phase transitions have been presented in Ref.~\cite{Pen2012}, which finds that kinetic-SZ data constrain bubble nucleation from false vacuum decay to happen very recently. We defer the studies of such inhomogeneities as well as the cosmological effects of neutrino self-interactions, (partial) annihilation, and dark radiation to future investigations.

\item Finally, we note that for simplicity we fixed the false vacuum energy density $V_0$ to the energy density required to generate the relic neutrino masses. However, a substantial amount of the false vacuum energy could also convert into dark radiation. In general, $V_0$ is a free parameter of the model~\cite{Dvali:2016uhn}, which opens up the possibility that $V_0$ could be identified with the observed dark energy density.\footnote{Ref.\ \cite{Dvali:2016uhn} already noticed a potential connection between the neutrino vacuum condensate and dark energy, due to the surprising numerical coincidence of the dark energy and neutrino mass scales, and because the neutrino condensate is inherently connected to a new low-energy gravitational scale, $\Lambda_G$. However, the model does not solve the cosmological constant problem since it cannot explain why other Standard Model vacuum contributions, such as the Higgs condensate, do not contribute to the cosmological constant.} In such a ``decaying dark energy'' scenario, our Universe recently became dark-radiation dominated, will soon enter a matter-dominated era, and will continue to expand at a decelerating rate (see\ e.g.\ Refs.\ \cite{Goldberg2000, Chacko2004,delaMacorra2007,Dutta2009,Pen2012,Abdalla2012, Krauss2013,Landim2016} for similar considerations). The redshift of dark energy decay is constrained by Type IA supernovae data to $z_{s}\lesssim 0.1$ at the $2\sigma$ level \cite{Dutta2009}. The dark radiation bosons would not yield directly observable cosmological effects, despite their huge abundance, due to strongly suppressed interactions with Standard Model particles. However, they might yield observable signatures in non-cosmological contexts (see Refs.~\cite{Dvali:2016uhn,Dvali:2016eay}).
\end{itemize}

\section*{Acknowledgements}
We thank Shahab Joudaki for useful discussions and for helping with the KiDS CosmoMC. We thank Gia Dvali, Georg Raffelt, Edoardo Vitagliano, Cyril Lagger, Ue-Li Pen, Silvia Pascoli, C\'{e}line B{\oe}hm, and Sophie Marie Koksbang for valuable discussions, and Martina Gerbino for comments on the draft. CSL is supported by a Clarendon Scholarship and acknowledges support from Pembroke College, Oxford. LF thanks the International Max Planck Research School on Elementary Particle Physics for support while this work was begun. This research was supported in part by Perimeter Institute for Theoretical Physics. Research at Perimeter Institute is supported by the Government of Canada through the Department of Innovation, Science and Economic Development and by the Province of Ontario through the Ministry of Research, Innovation and Science. EC is supported by a Science and Technology Facilities Council Ernest Rutherford Fellowship, grant reference ST/M004856/2. SH is supported by a grant from the Villum Foundation.



\bibliographystyle{apsrev4-1}
\bibliography{ref}

\appendix
\onecolumngrid

\end{document}